# Giant and Continuous Ionic Current Oscillation Induced by Dynamic Surface Charge Regulation in Cylindrical Mesopores

Hongwen Zhang,[1, 2#] Yujie Zhao,[1, 2#] Zekun Gong,[1, 2] Chih-Yuan Lin,[3] Tianyi Sui,[4] Zuzanna S. Siwy,[3] and Yinghua Qiu[1, 2]*

1. Key Laboratory of High Efficiency and Clean Mechanical Manufacture of the Ministry of Education, State Key Laboratory of Advanced Equipment and Technology for Metal Forming, School of Mechanical Engineering, Shandong University, Jinan, 250061, China

2. Shenzhen Research Institute of Shandong University, Shenzhen, 518000, China

3. Department of Physics and Astronomy, University of California, Irvine, California, 92697, United States

4. School of Mechanical Engineering, Tianjin University, Tianjin, 300072, China

# H.Z. and Y.Z. contributed equally.

*Corresponding author: yinghua.qiu@sdu.edu.cn

## Abstract

Nanofluidic ionic oscillators based on the dynamic regulation of surface charges hold great promise for neuromorphic computing, biosensing, and ionic circuits. Here, by dynamically adjusting the local charge inversion on pore walls, we present a simple and effective strategy to achieve periodic current oscillations by harnessing the transient adsorption and desorption of $Ca^{2+}$ ions in cylindrical mesopores under concentration gradients. Based on the combined precision current measurements and multiphysics simulations, we demonstrate that local overadsorption of $Ca^{2+}$ ions may induce asymmetric (bipolar) charge distributions along the pore axis, which periodically reverses the direction of electroosmotic flow and modulates the local ion concentration inside the pore, generating highly regular current oscillations. Notably, both the oscillation frequency and the open-state probability of the pore vary nearly linearly with the applied voltage. Moreover, under dynamic voltage scanning, the system exhibits typical memristive hysteresis, and the switching between the "open" and "closed" states is highly reproducible. This work not only reveals the dynamic, heterogeneous surface charge regulation by divalent ions, but also provides a simple, material agnostic method for constructing ionic oscillators and memristors based on dynamic adsorption/desorption of multivalent ions.

## Introduction

Solid-state nanopores serve as a versatile platform for functional nanofluidic devices.[1, 2, 3] By leveraging diverse advanced micro/nanofabrication techniques, solid-state nanopores with controllable dimensions and tunable shapes have been fabricated in various membranes.[1, 4, 5] Besides, different materials endow distinct inner wall characteristics, and the properties of nanopore walls can be readily modified through physical or chemical functionalization.[6] These functional nanopores are widely applied in areas such as micro/nanofluidic sensing,[7, 8] detection of biological and chemical molecules,[1, 9] ionic amplifiers and memristors,[10, 11] and high-performance energy harvesting.[12, 13]

The highly confined fluid inside a nanopore is significantly influenced by the interfacial properties of pore walls, such as the wettability and surface charges.[2, 14, 15] The wall characteristics of the nanopore play a critical role in regulating the transport of fluid and ions through the pore. At the solid-liquid interface, surface charges generated by the ionization of surface groups electrostatically attract counterions in the solution, inducing the accumulation of counterions near the solid surface to form electric double layers (EDLs).[16] The presence of EDLs not only endows the nanopore with selectivity toward counterions[17] but also provides a migration pathway for their rapid transport through the pore.[18, 19] Leveraging the hydration properties of ions, when an external electric field is applied, the directional migration of counterions inside EDLs drives the liquid flow through the pore, generating electroosmotic flow (EOF).[20] EOF draws the solution from the pore entrance into the pore.[21, 22] When a concentration gradient exists across the pore, EOF modulates the ionic distribution inside the pore,[23, 24] leading to asymmetry in the ionic current amplitudes under opposite voltage polarities, a phenomenon known as ionic current rectification (ICR).[25, 26, 27] Through EOF-induced

ICR, the surface charge properties can be indirectly characterized by analyzing the ICR features, due to the dependence of the ICR phenomenon on the polarity and magnitude of the effective surface charge on the pore wall.[23] Here, by tuning the surface charge properties of the pore walls, the modulation of ionic and liquid transport inside the pore is achieved.

Charge inversion induced by multivalent ions is an important mechanism for regulating the charge property of pore walls.[28, 29, 30] Inside EDLs, the accumulation of counterions near a charged wall is modulated by the valence and concentration of ions in the solution. When multivalent counterions are present in the solution, and their concentration exceeds a certain critical threshold, they become excessively adsorbed onto the charged surface. In this case, the net electrical potential of the solid/liquid interface inverts in polarity.[28] As a result, coions in the solution act as the counterions after charge inversion, thereby reversing the direction of EOF. For EOF-induced ICR under a concentration gradient, the expected ICR direction is reversed upon surface charge inversion.

Charge inversion at charged surfaces has been observed and widely reported in various systems, demonstrating significant application value in areas such as drug delivery and DNA sequencing.[28] In this context, we find that the critical solution concentration for surface charge inversion is crucial for effectively controlling the occurrence and extent of this phenomenon. Different studies have reported varied threshold concentrations for surface charge inversion on different materials. Chou et al.[29] found that divalent $Ru(bpy)_3^{2+}$ induced charge inversion in fused silica nanochannels at concentrations as low as 0.1 mM. In silica and carboxylated silicon nitride nanochannels,[31, 32] the threshold for $Ca^{2+}$-induced charge inversion was approximately 10 mM. Using streaming current measurements, Van der Heyden et al.[33] reported that in $SiO_2$ nanochannels, the thresholds for $Ca^{2+}$ and $Mg^{2+}$ were as high as 350-400 mM, whereas the threshold for trivalent $CoSep^{3+}$ was as low as about 75-100 μM. The occurrence of charge inversion depends not only on the valence and concentration of ions but also on the surface properties of pore walls, electrolyte species, and experimental conditions.[28, 34] According to the strongly correlated liquid theory, strong electrostatic interactions among multivalent counterions are the core cause of charge inversion.[28] However, this theory can provide only an estimate of the threshold concentration necessary for charge inversion in a given system.

In the phenomenon of EOF-induced ICR under a concentration gradient, if one side of the solution contains multivalent ions that can cause charge inversion on the pore wall, when these ions are driven into the pore by EOF and they undergo dynamic adsorption and desorption on the pore wall. Consequently, the surface charge polarity of the pore will switch, thereby leading to a dynamic change in the direction of EOF, and oscillations in the ionic current.

The occurrence of current oscillations arises from time-dependent electrical conductivity in confined spaces, such as variations in ion concentration[2] or changes in resistance due to the particle translocation through the pore.[35] The current oscillation phenomena in pores also involve a variety of driving mechanisms, among which the transient blockage by ions or particles is the most direct cause. For example, Lee et al.[36] observed that random cation trapping in single-walled carbon nanotubes led to instantaneous current drops. On this basis, more regular oscillations can be generated when the blocking species can form and dissolve periodically in a conically shaped nanopore.[37, 38, 39] In solutions containing divalent cations, a negative voltage drives the cations to accumulate at the tip of the pore and form insoluble nanoprecipitates with anions. The current recovers after the precipitates dissolve, and new precipitates can form. Such cycles of precipitate formation and clearing produce oscillations whose frequency can be tuned by adjusting the voltage, ion concentration, and ion type. In micro-/nanopores, the wall properties can modulate the solution inside the pore, causing the ionic conductivity to change over time. The periodic current instabilities induced by the cyclic switching of surface charge polarity are referred to as ionic current oscillation,[40, 41] which underpins the construction of ionic oscillators and stochastic sensors, and holds broad application prospects in biomimetic ion channels,[42] biosensing,[37] and neuromorphic computing.[43] Coupling a chemical oscillator that generates periodic pH fluctuations with a pH-responsive nanopore can also achieve autonomous and continuous ionic current oscillations.[40, 44] Xiong et al.[45] used a poly(imidazolium)-modified nanopore to produce periodic current signals through the feedback loop of anion recognition/desorption and reversal of EOF direction, successfully mimicking the spike-encoding function of neurons and demonstrating promising potential in biomimetic information processing. Without any additional modification of pore walls, Siwy et al.[46] employed conical poly(ethylene terephthalate) (PET) nanopores to generate millisecond-scale current oscillations relying solely on transient adsorption and desorption of $Ca^{2+}$ ions on charged surfaces to achieve periodic switching of the surface charge polarity between positive and negative. Their system behaves similarly to a single-junction transistor and provides an experimental basis for developing ionic oscillators and stochastic sensors that require no chemical modification.

Compared with the oscillation mechanisms described above that rely on precipitation, chemical reactions, or complex modifications, the direct regulation of ionic current through dynamic changes in the charge property of pore walls is more straightforward. However, a systematic understanding of the periodic current oscillations induced by purely electrostatic adsorption of multivalent ions, which involves dynamic and non-uniform surface charge regulation, is still lacking. Therefore, in this work, we investigate the dynamic regulation of the surface charge on pore walls by $Ca^{2+}$ ions and the induced periodic current oscillations. Based on precision current measurements under concentration gradients across the pore, combined with multiphysics simulations and theoretical analysis, we demonstrate that the dynamic adsorption and desorption of $Ca^{2+}$ ions on charged pore walls can induce local charge inversion, thereby creating a bipolar charge distribution on pore walls. This dynamic charge distribution periodically reverses the EOF direction, altering the ionic concentration inside the pore. Highly regular periodic current oscillations are captured in the current-time traces. Experimental results show that both the oscillation event frequency and the “open” state fraction of the pore vary nearly linearly with the applied voltage, and the system exhibits typical memristive behaviors under dynamic voltage scanning. This study not only provides experimental and theoretical insights into the dynamic, non-uniform surface charge regulation induced by multivalent ions, but also lays a foundation for developing facile and controllable ionic oscillators based on electrostatic adsorption of multivalent ions. Moreover, the mechanism of ion current oscillations is applicable to mesopores that are more accessible than precision nanopores.

## Results

Electroosmosis-induced switching of ICR under $CaCl_2$ gradients.

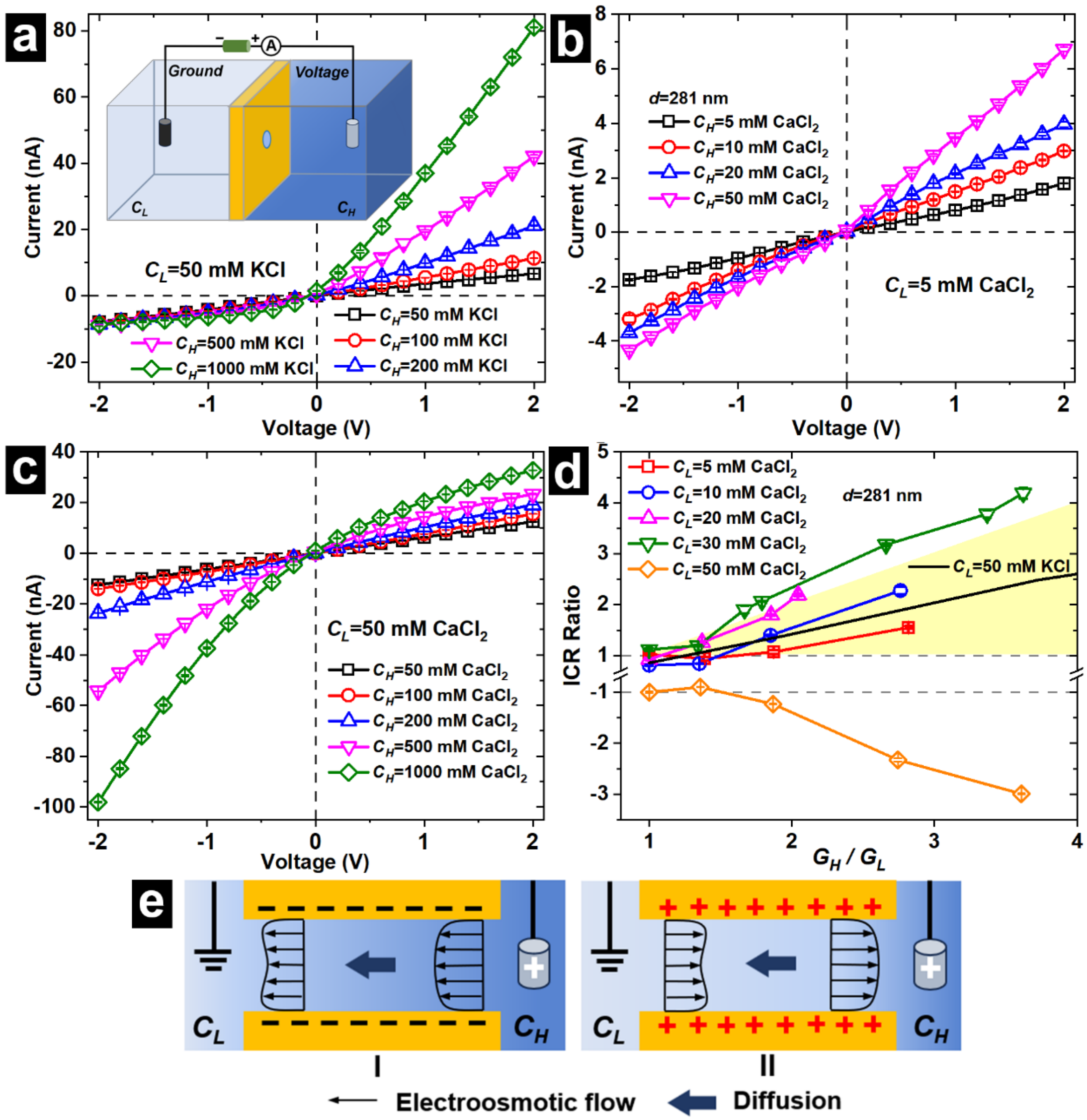

Fig. 1 **ICR under various salt gradients of KCl and $CaCl_2$ solutions.** (a) current-voltage (I-V) curves under different KCl concentration gradients. The inset shows the schematic diagram of the experimental setup. The ground and working electrodes are placed in the low- and high-concentration sides, respectively. (b-c) I-V curves under different $CaCl_2$ concentration gradients, with the low concentration equal to 5 mM (b) and 50 mM (c). (d) Variation of the ICR ratio with the solution conductivity ratio across the pore. The yellow area corresponds to the regime where the ratio of rectification degree and the ratio of conductivity of both solutions is less than 1. The diameter and length of the PET pore were ~281 nm and ~12 μm, respectively. (e) Schematic of the EOF direction and ion concentration distribution inside the pore with negative and positive surface charges.

The schematic diagram of the experimental setup is presented in the inset of Fig. 1a. Poly(ethylene terephthalate) (PET) membranes with a single cylindrically shaped mesopore are located between two aqueous reservoirs, filled with solutions of a low ($C_L$) and high concentration ($C_H$), respectively. During our nanofluidic experiments, KCl and $CaCl_2$ were considered as the monovalent and divalent electrolytes. Electrochemical characteristics of ionic current, including the current-voltage (I-V) and current-time (I-t) curves, were obtained under various concentration gradients, with the ground and working electrodes immersed in the low- and high-concentration solutions, respectively.

Fig. 1a plots the I-V curves obtained under concentration gradients of KCl solutions varying from 1 to 20, with the low concentration kept at 50 mM. The recorded I-V curves are non-linear and rectifying,[21, 22] such that different current values are recorded at equal voltage magnitudes but of opposite polarities. The EOF-induced ICR phenomenon results from the difference in ion concentration inside the pore under opposite biases, originating from the concentration gradient across the pore. The ICR ratio, used to quantitatively evaluate the degree of current rectification through mesopores, is calculated as the absolute value of the ionic current ratio at ±2 V, i.e., ICR ratio = $|I_{+2V}/I_{-2V}|$.[47, 48] Note that in the case with 50 mM KCl on both sides of the membrane, the I-V curves are symmetric, and the ICR ratio should be equal to 1. In our experiment, due to possible shape imperfections of the PET pores or minor differences in the solution level on both sides of the membrane, the obtained ICR ratio is ~0.94, very close to 1.

In the considered cases with $C_L$ = 50 mM and various values of $C_H$ of KCl solutions, the current values at negative voltages in the I-V curves remain almost unchanged, while the current values at positive voltages exhibit a gradual increase. This behaviour corresponds to the increased electrolyte concentration dragged into the mesopore by EOF. With $C_H$ varying from 100 to 1000 mM, the current value at positive voltages enhances from 11.35 to 81.11 nA, producing ICR ratios varying from ~1.4 to ~8.5.

From previous reports,[19, 24] the EOF-induced ICR ratio in nanopores can be predicted with the ratio of system resistances under opposite polarities. Considering the activity of the solutions, the conductivity ratio of KCl solutions across the pore is applied instead of the concentration gradient.[22] For KCl solutions with varying concentration gradients, the measured ICR ratio exhibits a linear increase as the conductivity ratio increases (Fig. 1d). Leveraging the precise detection platform provided by the pore, the EOF-induced ICR ratio can be used to evaluate the charge properties of the pore wall, as shown in our previous work.[23]

At solid-liquid interfaces, multivalent counterions that aggregate inside the EDLs can induce overscreening of surface charges, i.e., the appearance of charge inversion.[28] Similar to the investigation in previous studies,[29] an originally negatively charged surface exhibits a positive surface potential after charge inversion. Here, based on the direction dependence of the EOF-induced ICR on the sign of the surface potential, we extended the pore platform to investigate the regulation of surface charge properties by divalent $Ca^{2+}$ ions, by setting $CaCl_2$ concentration gradients across a mesopore. As described below, $CaCl_2$ gradients can induce not only charge inversion and corresponding inversion of ICR, but also ionic current oscillation and memristive properties both in I-t and I-V recordings.

I-V curves obtained under concentration gradients of $CaCl_2$ ($C_H/C_L$) varying from 1 to 10 are shown in Fig. 1b, with the low-concentration fixed at 5 mM. Please note that in our experiments, the high concentration can be changed from 5 to 1000 mM. Due to the appearance of the dynamic regulation of surface charges by $Ca^{2+}$ ions,[32] I-V curves present instabilities under larger salt gradients (see below) that prevent us from reporting ICR ratios. In the case without a salt gradient, i.e., with 5 mM $CaCl_2$ on both sides, the I-V curve is linear, as expected. As the concentration gradient is introduced by increasing $C_H$ from 5 to 50 mM, the current under positive voltages becomes larger than that under negative voltages. In these cases, due to the relatively low concentration of $Ca^{2+}$ ions (less than 50 mM), the pore surfaces remain negatively charged. Supplementary Fig. 1 shows the I-V curves under various salt gradients with $C_L$ = 10, 20, and 30 mM.

The calculated ICR ratios of the stable I-V curves are summarized in Fig. 1d. For cases of $CaCl_2$ solutions with $C_L$ not higher than 50 mM, the ICR ratios share a similar trend to those in KCl solutions, i.e., a linear increase with the solution conductivity ratio. However, due to the different values of $C_L$, the increasing slopes are different. At $C_H/C_L$ = 4, the ICR ratio rises from ~1.2 to ~2.1 as $C_L$ changes from 5 to 30 mM, increasing by 75%. As the concentration on the high-concentration side increases, the diffusion effect is enhanced at positive voltages due to the enlarged concentration difference ($C_H-C_L$), leading to an increase in ICR ratios. Similar dependence of ICR ratios on the ratio of the solution conductivity was also observed with another 484-nm-in diameter PET pore. (Supplementary Fig. 2)

From Supplementary Fig. 3, when $C_H$ reaches 100 mM, while keeping $C_L$ =5 mM, the I-V curves start to show a pinched hysteresis as the voltage is scanned in both the forward and reverse directions. We hypothesize that at these conditions charge inversion appears (see below) and affects transport properties of the pore. As plotted in Fig. 1c, for the I-V curves recorded in $CaCl_2$ solutions with $C_L$ = 50 mM, with salt gradients varying from 1 to 20, opposite ICR is obtained for all voltages, i.e., the current under positive voltages is smaller than that under negative voltages. Considering that the ICR is EOF-induced, the switch in the ICR direction indicates that the effective surface potential indeed changes from negative to positive, as predicted by the classical Helmholtz-Smoluchowski equation Eq. 1.[14]

$$v_{EOF} = -\frac{\varepsilon\varphi V}{\eta L} \quad (1)$$

where $v_{EOF}$, $\varepsilon$, $\varphi$, and $\eta$ are the velocity of induced EOF, dielectric constant, surface potential, and solution viscosity, respectively. $V$ and $L$ are the applied voltage and the pore length.

The reversed ICR suggests that a substantial influx of $Ca^{2+}$ ions into the mesopore leads to the charge inversion through over-adsorption of divalent $Ca^{2+}$ ions onto the charged surfaces. Consequently, the effective surface charge of pore walls shifts from negative to positive. As a result, $Cl^-$ ions act as counterions, and the direction of EOF is in the opposite direction to the electromigration direction of $Ca^{2+}$ ions. At positive voltages, the EOF points toward the high-concentration solution, drawing the low-concentration solution into the pore, resulting in a smaller current.

Based on our experimental data, we can conclude that once the concentration of $CaCl_2$ exceeds a threshold value,[28] the occurrence of the overscreening to surface charges can result in charge inversion. Considering the reversed ICR, the ICR ratio was calculated by ICR ratio = $-|I_{-2V}/I_{+2V}|$. As the ratio of the solution conductivity increases, the ICR ratio also exhibits an increasing trend (orange curve, Fig. 1d).

Charge inversion via $Ca^{2+}$ ion adsorption.

Following the strongly correlated liquid (SCL) theory,[28] the charge inversion originates from the strong correlation between multivalent counterions and bare surface charges at solid-liquid interfaces. The prerequisite for the charge inversion is that the concentration of multivalent ions in the solution must exceed a certain threshold, i.e., the critical concentration, $C_T$. According to the SCL theory, the threshold concentration can be evaluated using Eqs. 2-4.[28, 32]

$$c_T = \left| \frac{\sigma_b}{2 r_{ion} Z e} \right| \exp\left( \frac{\mu_c}{k_B T} \right) \tag{2}$$

where $\sigma_b$, $e$, $k_B$, and $T$ are the bare surface charge density, elementary charge, Boltzmann constant, and temperature, respectively. $r_{ion}$, and $Z$ are the radius and valence of the adsorbed multivalent ions, respectively. $\mu_c$ is the chemical potential due to correlation effects, described by Eq. 3.

$$\mu_c = -k_B T \left( 1.65\Gamma - 2.61\Gamma^{1/4} + 0.26 \ln \Gamma + 1.95 \right) \tag{3}$$

where $\Gamma$ is the coulomb coupling parameter, representing the interaction strength among multivalent ions inside the Stern layer, calculated using Eq. 4.

$$\Gamma = \frac{1}{4 k_B T \varepsilon_0 \varepsilon_r} \sqrt{\left| \frac{e^3 Z^3 \sigma_b}{\pi} \right|} \tag{4}$$

where $\varepsilon_0$, $\varepsilon_r$ are the vacuum permittivity and the relative permittivity, respectively.

Based on the SCL theory, through the solution of coupled Eqs. 2-4, the threshold concentration for the appearance of charge inversion in $CaCl_2$ solutions under different surface charge densities is obtained, as shown in Supplementary Fig. 4. Here, the parameters were chosen as: $r_{ion}$ = 0.4 nm, T=298 K.[32] By assuming $\sigma_b$ = 40 mC/m$^2$, the $C_T$ is predicted as ~37 mM (Supplementary Fig. 4).

Under concentration gradients across the pore, the uniform charge inversion of the pore can be sustained only when the low concentration reaches $C_T$. By analyzing the EOF-induced ICR phenomena, we attempted to determine the threshold concentration of $CaCl_2$ in PET mesopores, with the variation of $C_L$ from 5 to 50 mM. As shown in Figs. 1c and 1d, charge inversion appears when $C_L$ reaches 50 mM, which is the estimated $C_T$ in our system.[33]

The phenomenon of charge inversion had also been reported for nano/micropores exposed to a range of multivalent electrolytes.[30, 32] The modulation of the surface charge property by multivalent counterions was detected as a change in the conductance. While previous research mainly focused on the uniform charge inversion that happened on the pore walls, here, taking advantage of polymer mesopores with a large length of ~12 μm, the dynamic modulation process of surface charges by $Ca^{2+}$ ions can be captured.[29] In the considered scenarios, charge inversion may appear on a portion of the pore wall, leading to the formation of dynamic surface charge patterns.

## Ionic current oscillation via surface charge modulation and dynamic direction of EOF.

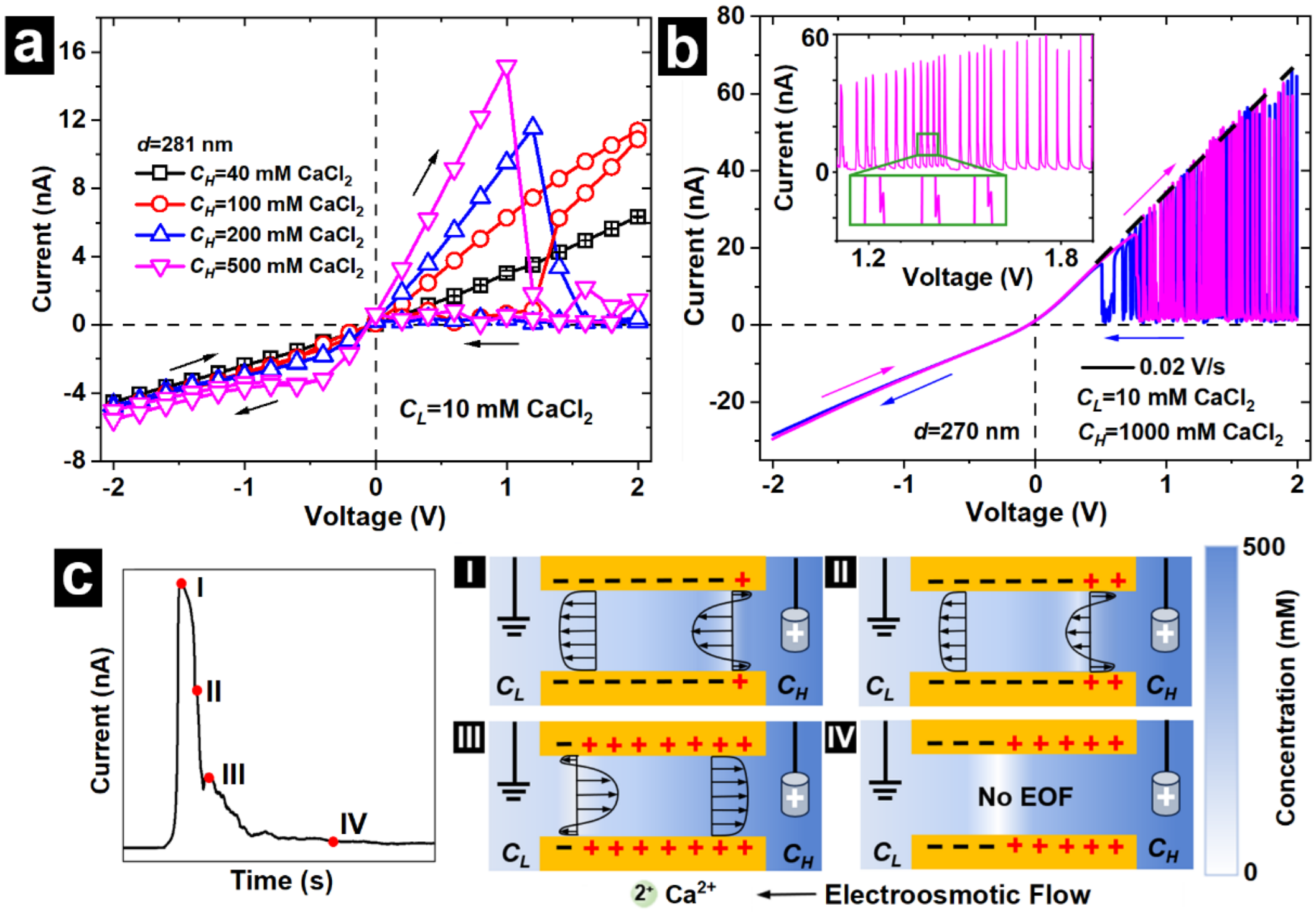

Fig. 2 **Ion current hysteresis and instabilities under different $CaCl_2$ concentration gradients with local charge inversion on pore walls**. (a) I-V curves recorded by Keithley 6487 with $C_L$= 10 mM $CaCl_2$ and $C_H$ varying from 40 to 500 mM $CaCl_2$. Black arrows show the scanning direction of the voltage. (b) I-V curves recorded by CHI 760E with $C_L$= 10 mM $CaCl_2$ and $C_H$=1000 mM $CaCl_2$. Arrows indicate the scanning direction of the applied voltage from −2 V to 2 V. The scan rate is 0.02 V/s. The inset shows the magnified view of the current oscillation section under positive voltages. (c) Event example and corresponding schemes of the solution distribution and electroosmotic flow inside the pore at stages I-IV. Positive surface charges are induced by the charge inversion due to the adsorption of $Ca^{2+}$ ions.

The I-V curves in Fig. 2a were recorded for a 281 nm in diameter mesopore, at $C_L$ = 10 mM $CaCl_2$ and gradually increasing $C_H$ to 100 mM $CaCl_2$ (red curve). The recorded I-V curve exhibits clear hysteresis,[48, 49, 50] i.e., current values at the same voltage differ during forward and reverse voltage scans. During the forward scan from 0 to 2 V, the current increases continuously and remains higher than the corresponding current under negative voltages. However, in the reverse scan from 2 V back to 0 V, the current drops rapidly to almost zero at ~1.4 V, and remains at the low value until reaching ~0 V (red curve of Fig. 2a). The ultra-low current indicates that the pore stays at a “closed” state. This phenomenon is robust and appears for different values of $C_L$, i.e., 5 and 20 mM (Supplementary Fig. 3). When $C_H$ reaches 200 and 500 mM, the higher ion concentration induces an unexpected decrease in the current at ~1.2 and ~1 V during the forward voltage scan from 0 to 2 V (see blue and pink curves in Fig. 2a). This negative differential resistance (NDR)[51, 52, 53] can serve as an indicator of the dynamic change in surface charge associated with charge inversion.

We hypothesize that the occurrence of the ultra-low current originates from the dynamic modulation of the surface charge properties by the adsorption of $Ca^{2+}$ ions, which may induce a bipolar charge distribution of the pore walls.[47, 54] Under positive voltages, $Ca^{2+}$ ions migrate from the high-concentration solution and modulate the charged surfaces to induce charge inversion. This results in a portion of the pore walls becoming positively charged, a significant departure from the case with uniform charge inversion stated above (Fig. 1). Due to the concentration gradient across the pore, the combined migration and diffusion facilitate the movement of $Ca^{2+}$ ions through the pore.[23] Consequently, the concentration of $Ca^{2+}$ ions varies along the pore axis from a high concentration, which is above $C_T$ and can induce charge inversion, to a low value, which cannot trigger the charge inversion. In this case, the pore walls with originally uniform negative charge transform into a bipolar charge distribution, with both negatively and positively charged regions on the low- and high-concentration sides, respectively.

We would like to note that a unipolar charge distribution also may be possible, with the neutral and negatively charged regions (Supplementary Fig. 5).[47] In this case, for the neutral region, the bare negative charges are exactly compensated by the adsorbed $Ca^{2+}$ ions.

During the reverse voltage scan from 2 to 0 V, due to the increased time with the applied voltage across the pore, a larger area of the pore wall becomes charge-inverted. In the I-V curve at $C_H$ = 500 mM, as the negative voltage changes further from −0.2 to −1 V, the current displays a plateau or a decreasing trend (pink curve of Fig. 2a and Supplementary Fig. 3). This NDR phenomenon arises from the switching of the salt concentration from a high to a low value, denoting the disappearance of the charge inversion on pore walls. In this case, at low voltages, the migration speed of counterions remains slow, producing a weak EOF. Due to the strong ion diffusion from the high- to the low-concentration side, the pore surface can maintain charge inversion. However, when the voltage reaches ~−0.6 V, the charge inversion on the pore walls starts to disappear, and the EOF from the low- to high-concentration side fills the pore with the diluted $CaCl_2$ solution, which produces a shallower slope in the I-V curve. The disappearance of the charge inversion may be attributed to the fact that the diffusive amount of $Ca^{2+}$ ions from the high- to the low-concentration side cannot maintain the threshold concentration of charge inversion, when $Ca^{2+}$ ions electromigrate in the opposite direction, i.e., from the low- to the high-concentration side.

Supplementary Fig. 6 shows example I-V curves that exhibit ionic current fluctuation during the voltage scan at positive voltages, i.e., at the “closed” state, which deserves detailed investigation at a high sampling frequency. To investigate the detailed characteristics of these current fluctuations, a larger concentration gradient was adopted with 10 mM : 1000 mM $CaCl_2$ solutions, because a higher salt concentration facilitates the appearance of the charge inversion. Multiple I-V scans were performed at a sampling frequency of 1 kHz using a CHI 760E electrochemical workstation. Fig. 2b plots the I-V curve obtained at a voltage scan rate of 0.02 V/s. For negative voltages and positive voltages below ~0.5 V, the stable current presents a clear ICR phenomenon without the occurrence of charge inversion. As the voltage increases further from ~0.6 to 2 V in the forward direction, distinct current fluctuations appear. As shown in the inset of Fig. 2b, the current fluctuations include multiple periodic oscillations. Please note that due to the presence of high-frequency current oscillations under positive voltages in both scanning directions, to provide a clear presentation of the current oscillations, pink and blue colors are used to show the forward and backward voltage scans. Supplementary Fig. 7 shows the curve of the second scanning loop. In the I-V recordings from ~0.6 to 2 V, the current clearly oscillates between a high value (“open” state) and a low value (“closed” state). The high values in the “open” state are directly proportional to the applied voltage, as depicted by the black dashed line, while the low current remains largely voltage independent. Considering the dynamic modulation of surface charges by $Ca^{2+}$ ions, we hypothesize that the “open” state of the pore corresponds to the pore with uniform negative surface charges, whereas the “closed” state results from the bipolar surface charge distribution. [54] Even though the current fluctuations might be reminiscent of current oscillations induced in conically shaped nanopores by precipitation of weakly soluble salts, the mechanism of oscillations in our system is completely different. The opening of pores here is significantly larger than the pores where precipitates were induced. Our pores are symmetric in shape, and all salts used are highly soluble.[55]

Fig. 2c presents a typical event of an ionic current pulse, and the corresponding schemes that link current instabilities to the dynamic modulation of surface charges by $Ca^{2+}$ ions. The fluctuation event starts from an “open” current and sharply decreases to a local minimum value. This is followed by a small increase in the current and a gradual reduction of the current to the “closed” value. Note that the smaller peak near the main current peak appears in all the events (see below). Under a $CaCl_2$ concentration gradient, the varying concentration of $Ca^{2+}$ ions along the pore axis enables the pore to have different local charge properties, which can also be adjusted by the applied voltage. Note that we focus our analysis on $C_L$ below $C_T$, i.e., below 50 mM $CaCl_2$, while with $C_H > C_T$, charge modulation can happen on the pore surfaces connected to the high-concentration region (I, Fig. 2c). Under positive voltages, EOF drags the high-concentration $CaCl_2$ into the pore continuously. The charge modulation on the pore surface expands fast (II, Fig. 2c), until the pore surface almost fully becomes charge-inverted. When a large fraction of the pore walls becomes positively charged, the EOF reverses and becomes directed from the low- to the high-concentration side (III, Fig. 2c). The reversed EOF brings a diluted solution into the pore,[21, 22] which facilitates desorption of $Ca^{2+}$ ions from the pore wall near the low-concentration side. As a result, the pore walls are rendered with a surface charge distribution characteristic of a bipolar diode. Due to the continuous desorption of $Ca^{2+}$ ions, the modulation area can reach a critical value which produces a very low current at the “closed” state (IV, Fig. 2c). Please note that at the “closed” state of the pore, the EOF inside the bipolar pore is extremely weak.[54] Later, as $Ca^{2+}$ ions continuously desorb from the surface, the pore walls relieve the charge inversion and become negatively charged again.[29] The “open” state recovers, presenting an open current.

The current oscillations originate from the unstable binding of $Ca^{2+}$ ions to the charged surface, which can detach when the diluted solution is dragged into the pore by the reversed EOF. As the portion of the pore walls with charge inversion diminishes, the net EOF goes back to the positive direction, allowing $Ca^{2+}$ ions to re-enter the pore from the high-concentration side. This leads to continuous switching of the surface properties between uniformly negatively charged and bipolarly charged states, presenting oscillation in the current trace.

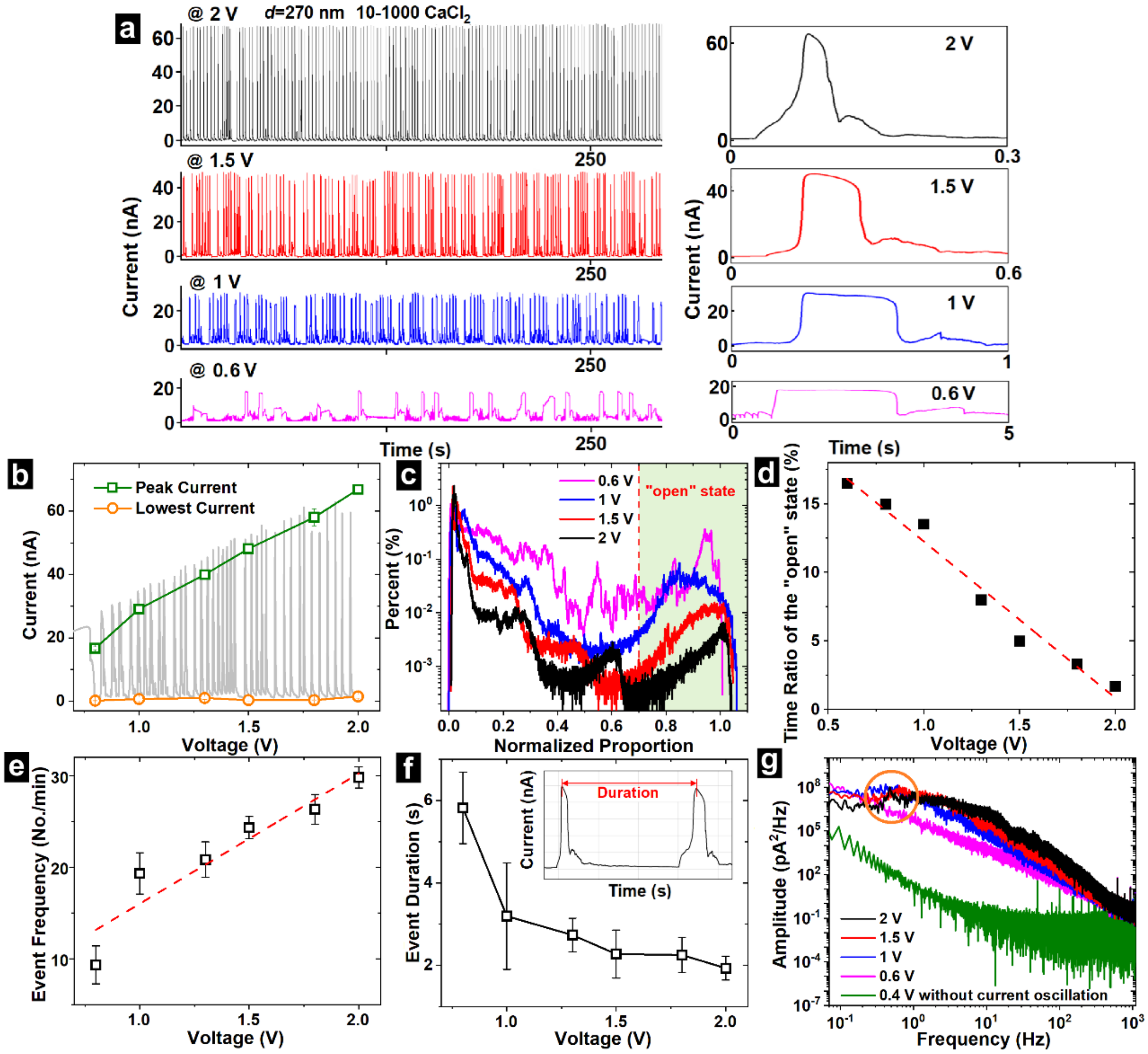

Fig. 3 **Characteristics of current oscillation in the I-t traces.** (a) I-t curves recorded by Axopatch under different voltages. The sampling frequency was 10 kHz. The insets show corresponding examples of oscillation events at varying voltages. (b) Statistics of the peak current corresponding to the “open” state and the baseline current corresponding to the “closed” state in the I-t curves. (c) Histogram of the normalized current values relative to the peak current in the I-t trace shown in (a). 0.02 nA was used

as the interval. The percentage is calculated by dividing the amount of each discrete current value ($N_I$) by the total sample number ($N_{total}$), i.e., Percent = ($N_I$ / $N_{total}$)×100%. (d) The fraction of time the pore remains open as a function of voltage. (e) The number of pore openings per minute as a function of voltage. (f) The event duration between adjacent spikes under different voltages, with an inset illustrating the measurement positions for the event duration. (g) Current power spectral density of the I-t traces under different voltages. The orange circle labels the current power spectral density contributed by the current oscillation in the I-t traces.

To capture the detailed dynamics during the modulation of charged surfaces by $Ca^{2+}$ ions, the periodic behaviour of current fluctuations at varying voltages is examined using an Axopatch 200B amplifier at a sampling frequency of 10 kHz. From the collected I-t traces shown in Fig. 3a, the current exhibits periodic oscillations, corroborating the phenomenon presented in Fig. 2b, i.e., the pore continuously switches between the “closed” and “open” states.

As the next step, the variation of current values and fluctuation behaviors are analyzed under different voltages. Fig. 3b summarizes the magnitude of current peaks (open states) and the minimum current values (closed states) in the events. Both values in the I-t traces share similar values to those shown in the I-V curves (the gray background curve). At 1.5 V, the peak current in the I-t trace is ~48.7 nA, differing by only ~4.3% from the value of ~50.8 nA recorded in the I-V curve. As the voltage increases, the peak magnitude, corresponding to the fully open current of the pore, presents a linear increase. Interestingly, the closed current stays at nearly a constant level of ~1 nA at different voltages. Note that the closed current values are much smaller than the corresponding currents at negative voltages, which also exhibit voltage dependence (Fig. 2b).

To further examine the properties of current oscillations, the normalized current histograms were calculated as Percent = ($N_I$ / $N_{total}$)×100%, in which $N_I$ and $N_{total}$ are the number of each discrete current value at intervals of 0.02 nA in the 500-second-long current trace, and the corresponding total number of current values, respectively. Note that under the same sampling frequency, the number of data points directly depends on the duration. From Supplementary Fig. 8, as the voltage increases, the peak current gradually rises, while its corresponding percentage gradually decreases.

Ionic current values in the traces are subsequently normalized to the open current, i.e., the averaged peak current (Fig. 3c). We define the “open” state of the pore when the current reaches above 70% of the peak value. Fig. 3d presents the dependence of the open pore probability on the applied voltage. It’s interesting to see that as the voltage increases, the open time portion of the pore decreases linearly. At a low voltage of 0.6 V, the pore’s open time accounts for ~16.5% of the total time, which decreases to ~2% at 2 V, representing a reduction of ~90%. This indicates that a larger voltage leads to a higher percentage of the current recording in the “closed” state of the pore.

Simultaneously, the oscillation frequency and the event duration are analyzed based on the recorded 6-minute-long traces. Fig. 3e shows that the oscillation frequency is directly proportional to the applied voltage. The application of 2 V enables ~30 events per minute, which is more than three times the event frequency of ~9 events per minute at 0.8 V. Fig. 3f plots the corresponding duration of the events, obtained by measuring the time intervals between adjacent peaks, as shown by the inset. The event duration also reflects the oscillation rate of the pore. At 1 V, the event duration is ~3.1 s, which decreases to ~1.9 s as the voltage increases to 2 V, by ~39%. This may be attributed to the enhanced migration of $Ca^{2+}$ ions and EOF under higher voltages, which accelerates the process of both the formation and removal of charge inversion. Additionally, we analyzed the full width at half maximum (FWHM)[56] of the current peak in the I-t traces, which shows the duration for the current at 50% of the peak value (Supplementary Fig. 9). At 1 V, the FWHM is ~352 ms, which decreases to ~49 ms at 2 V, by ~86%. The significant decrease in FWHM with increasing voltage corresponds to the shortened time in the "open" state of the pore.

The power spectral density of ionic current recordings (Fig. 3g) is obtained based on the I-t traces to show the current noise within the frequency domain of 0.07-1000 Hz.[37] At 0.4 V, due to the weak EOF inside the pore, no oscillation occurs in the current trace. The stable current signal indicates that the pore remains continuously at the "open" state. The current power spectral density exhibits a low-frequency 1/f noise and a medium-frequency white noise, consistent with observations in previous experimental studies, Fig. 3g.[57, 58] As the voltage increases to 0.6 V, ionic current oscillation appears, and the pore keeps switching between the open and closed states. In this case, the current power spectral density exhibits 1/f noise across the entire frequency band (pink curve), which we believe can originate from the dynamic current response under the modulation of the surface charge by $Ca^{2+}$ ions. During the switching process between the "open" and "closed" states of the pore, high-frequency small-amplitude current fluctuations appear due to the dynamic adsorption and desorption of the $Ca^{2+}$ ions on the pore wall. For instance, Supplementary Fig. 10a presents ~10 small-amplitude current fluctuations within 1 second at 2 V, when the pore stays in the closed state. This corresponds precisely to the characteristic peak of ~10 Hz at 2 V (orange circle in Fig. 3g).[37]

Note that the oscillation events in the current traces correspond to the portion at low frequencies from ~0.1 to 1 Hz, which is also modulated by the applied voltage.[45] Due to the less regular oscillations in the I-t trace at 0.6 V, no distinct characteristic peaks appear in the power spectrum. As the voltage increases to 2 V, the current oscillations, switching between “open” and “closed” states of the pore, become much more regular, and induce the appearance of a characteristic peak at ~0.5 Hz in the current power spectral density profile (Fig. 3g), which is highly consistent with the event duration of ~1.97 s between adjacent peaks.

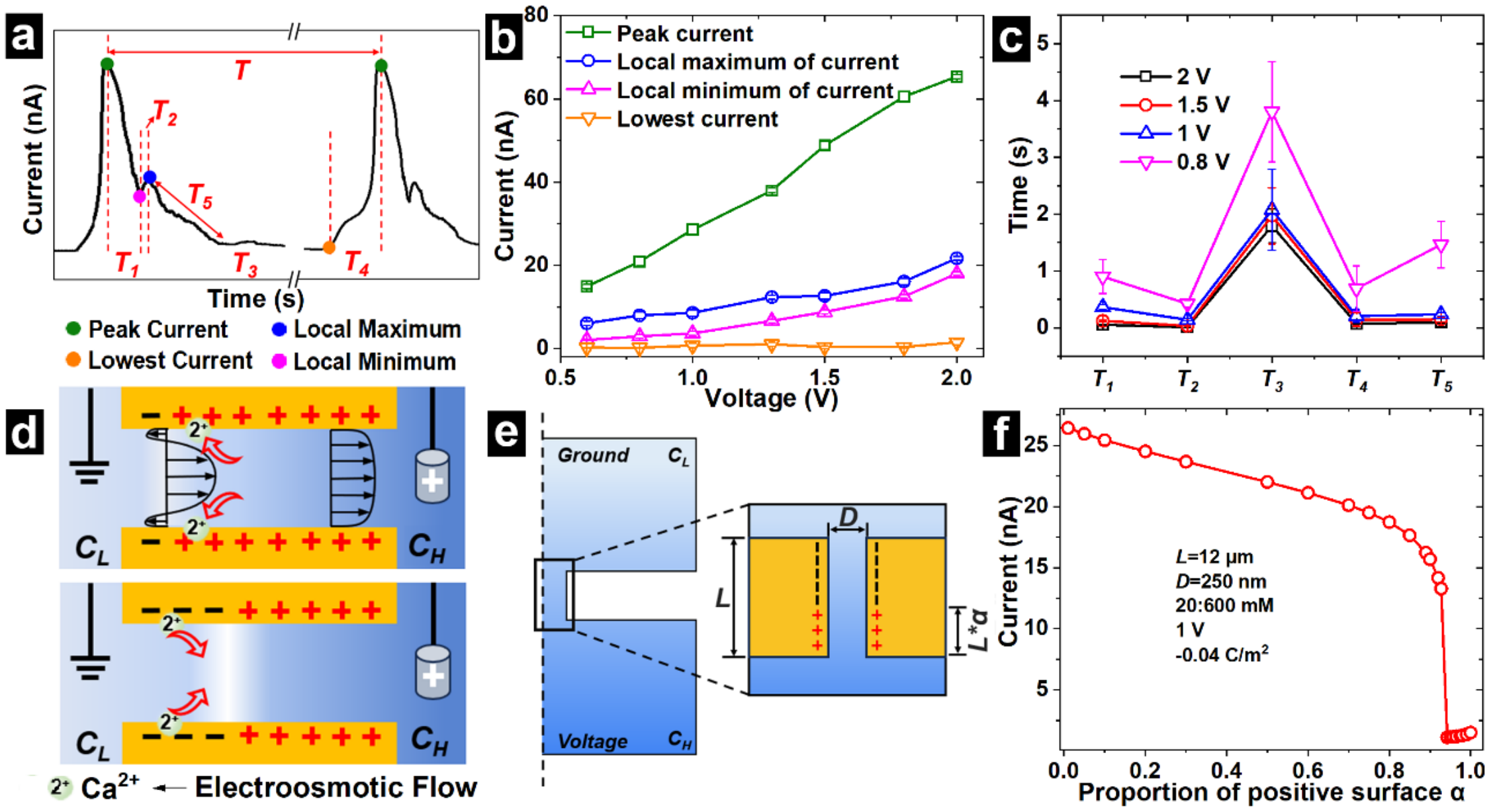

Fig. 4 **Analysis of individual oscillation events at different voltages**. (a) Schematic diagram of the event, including the current and time parameters. (b) Characteristic current values of each event. (c) Duration of each period within an oscillation event. (d) Schematic diagram of ion adsorption to pore walls and desorption from pore walls. (e) Schematic diagram of the simulation under concentration gradients. (f) Current variation in the mesopore with different proportions of positive surfaces (α).

As shown in the right-hand panel of Fig. 3a, oscillation events at different voltages exhibit a remarkably similar profile, including both the large-amplitude current peak and a small peak next to the main peak during the current-decreasing period. Below we analyze in detail how the events' characteristics inform on what happens in the pore.[45]

Based on the current features modulated by the dynamic adsorption and desorption of $Ca^{2+}$ ions,[32] in each oscillation event, four characteristic current states can be extracted from the current trace, i.e., the peak current ($I_{peak}$), local minimum ($I_{Min}$), local maximum ($I_{Max}$), and the lowest current ($I_{closed}$), as labeled by the different colored dots in Fig. 4a. The peak current denotes the open state of the pore. As $Ca^{2+}$ ions continuously enter the pore and adsorb on pore walls, the appearance of charge inversion results in a significant drop in current. With the area of charge inversion undergoing expansion, the bipolar pore presents the local current minimum ($I_{Min}$). With the further increase in the fraction of the pore walls with charge inversion, the direction of EOF reverses, and the current exhibits a small increase in the local current maximum ($I_{Max}$). However, reversed EOF drags diluted solution into the pore, facilitating the desorption of $Ca^{2+}$ ions from the surface (Fig. 4d) and inducing the current decrease due to the appearance of bipolar surface distribution on pore walls. At the closed state, EOF is extremely weak, and the solution concentration becomes more uniform. At the end of the event, a large amount of $Ca^{2+}$ ions desorb from the surface abruptly, and the pore recovers to the open state ($I_{peak}$). Please note that at high voltages, $I_{Min}$ is higher than the lowest current. This may be caused by the strong EOF that enables faster modulation of pore-wall characteristics by $Ca^{2+}$ ions. During this process, the formed bipolar distribution of pore walls at $I_{Min}$ deviates from the optimal bipolar configuration with maximum ion depletion inside the pore, which induces $I_{closed}$. In contrast, at a low voltage of 0.6 V, the weaker EOF allows a slower ionic modulation. In this case, $I_{Min}$ is very close to $I_{closed}$ (Fig. 4b). Fig. 4b summarizes the open and closed currents recorded at different voltages. Except for the lowest current $I_{closed}$, all other characteristic currents present a linear increase with the applied voltage.

In order to support the experimental findings and the qualitative hypothesis presented above, we developed a simulation model with coupled Poisson-Nernst-Planck and Navier-Stokes equations to explore the current variation with the changes in the distribution of surface charges. Fig. 4e shows the schematic diagram of the modeling. Detailed boundary conditions are listed in Supplementary Tab. 1. In the simulation, the pore length, diameter, and surface charge density were 12 μm, 250 nm, and ±0.04 $C/m^2$, corresponding to the experimental conditions. A concentration gradient of 600 mM : 20 mM KCl solution was set across the pore, because highly concentrated divalent ions cannot be solved with the classical mean-field equations. Fig. 4f shows that when the portion of the pore walls with positive surface charges, i.e., the charge-inverted portion, increases, the current through the pore first decreases to a current minimum and then increases to a value with the uniformly positive surface. Our simulation confirms the appearance of the local minimum when α ~0.94. We also investigated the current variation under different concentration gradients as the surface charge distribution varied, all of which exhibits a similar profile (Supplementary Fig. 11). Moreover, as the concentration gradient increases, the α value where the minimum current occurs increases.

Each current event, with a total duration of $T$, can be divided into 4 sections, with durations from $T_1$ to $T_4$, respectively, that denote the time for ion adsorption to pore walls, presence of the reversed EOF, ion desorption that leads to bipolar surfaces (almost no EOF), and abrupt ion desorption. For zone 3, during the ion desorption to form the bipolar surface, the current first decreases rapidly and then approaches the saturation slowly. We denote the duration for the fast desorption (Fig. 4d) as $T_5$. With the analysis of more than 50 events at each voltage, all durations are extracted and summarized in Fig. 4c. As the voltage increases, all durations decrease, indicating the current oscillation becomes faster. This corresponds to the fast dynamic modulation of surface charges by $Ca^{2+}$ ions due to faster EOF at higher voltages.

At low voltages (e.g., 0.8 V), when the current reaches its minimum, small-amplitude oscillations are more likely to occur, as shown in Supplementary Fig. 12. Because the weak EOF at low voltages can drag a small amount of diluted solution into the pore, some $Ca^{2+}$ ions undergo dynamic desorption from the wall and re-adsorption to the wall. Under these conditions, the EOF can only cause localized, minor disturbances to the existing dynamic equilibrium of positive and negative charge distribution on the wall, leading to periodic fluctuations in the current around a relatively low level.

## Memristive characteristics of pores subjected to $CaCl_2$ gradients and charge inversion.

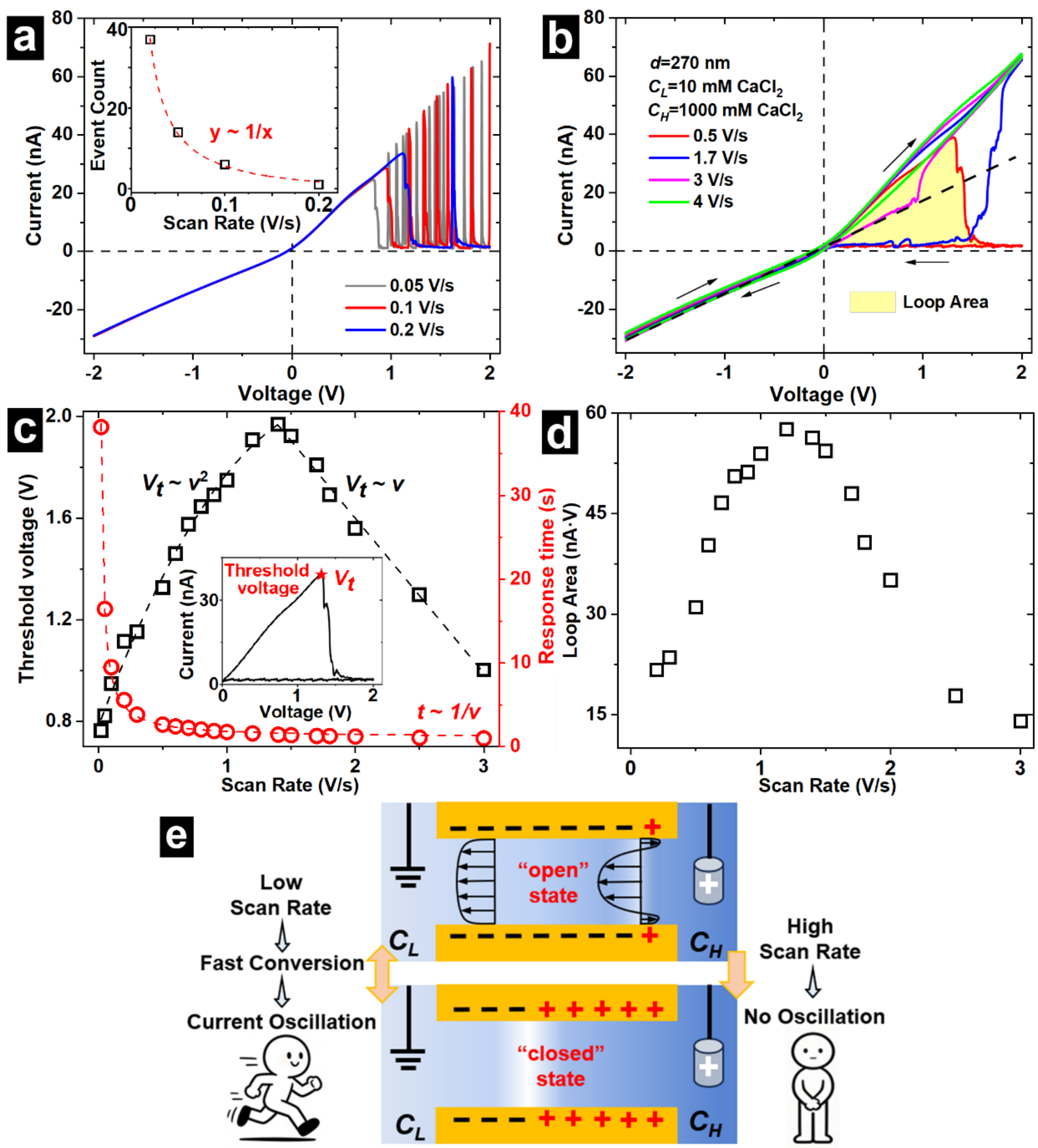

Fig. 5 **Dynamic current characteristics under different scan rates**. (a) I-V curves at scan rates varying from 0.05 to 0.2 V/s, where charge oscillation occurs at 0.8 to 2 V. The inset shows the count of oscillation events captured in the I-V curves. (b) I-V curves at different scan rates. (c) Threshold voltage $V_t$ for the appearance of charge inversion at different scan rates, corresponding to the voltage at which negative differential resistance (NDR) occurs. The red curve illustrates the response time $t$ for charge inversion, calculated as $t = V_t / v$, where $v$ is the scan rate. The inset illustrates the determination of $V_t$ for charge inversion. (d) Loop area enclosed by the I-V curve at different scan rates. (e) Schematic diagram of ion current oscillation.

Finally, we considered the interplay between the time required for ion transport within the pore and the adsorption/desorption processes at the pore surface, and probed whether our system exhibits memristive characteristics under dynamic voltage scanning.[48, 55]

As shown in Figs. 5a and 5b, the scan rate of voltage can modulate the shape and current oscillations of the I-V curves. At a low scan rate of 0.01 V/s, the I-V curve can capture high-frequency oscillations in the current (Fig. 5a). As the scan rate gradually increases to 0.2 V/s, the number of oscillation events observed during the forward scan from 0.02 to 0.2 V/s decreases from 37 to 1, and the oscillation frequency in the I-V curve gradually decreases (inset of Fig. 5a). This may be caused by the disturbance caused by the gradually increasing voltage on the adsorbed ions on the pore wall, making it more difficult to reach the bipolar current minimum, which corresponds to the steady state under the “closed” condition.

Fig. 5b shows that as the scan rate increases to 0.5 V/s or higher, clean current hysteresis loops appear in the I-V curves. Supplementary Fig. 13 shows the complete I-V curves for scan rates from 0.2 to 1 V/s (forward-scan charge inversion) and from 1.4 to 3 V/s (reverse-scan charge inversion). At 0.5 V/s, the current presents a sharp drop at ~1.3 V in the forward scan from 0 to 2 V, indicating the occurrence of the NDR phenomenon. This suggests that the concentration of $Ca^{2+}$ ions adsorbed on the pore wall has reached a threshold value, triggering the surface charge inversion. The voltage corresponding to the occurrence of NDR is defined as the threshold voltage $V_t$, indicated in the inset of Fig. 5c. Under different scan rates, $V_t$ exhibits an overall quadratic growth trend with the scan rate. The time taken for the voltage to increase from 0 to $V_t$ is defined as the response time $t$, calculated by $t = V_t / v$ (inset of Fig. 5c). The response time is inversely proportional to the scan rate, meaning that a higher scan rate induces a shorter response time. The corresponding amount of charge $Q$ passing through the pore when NDR occurs is obtained by integrating $Q = I * t$ (Supplementary Fig. 14).

Fig. 5b also demonstrates that as the scan rate rises from 0.5 to 1 V/s, the threshold voltage $V_t$ increases from 0.76 to 1.74 V. Although the threshold voltage for the charge inversion increases, the amount of charge passing through the pore remains essentially unchanged, remaining at ~50 nC. When the scan rate further increases to 1.4 V/s (Supplementary Fig. 13b), due to the short time for the application of voltage across the pore, even though no NDR phenomenon occurs during the forward scan, a clear NDR effect appears in the reverse scan, denoted by a current drop at ~1.8 V. For the scan rate from 1.4 to 3 V/s, the threshold voltage for charge inversion in the backward scan presents a linear decreasing trend (Fig. 5c). Interestingly, at the scan rate of 2.5 and 3 V/s, the pore does not show the closed state, instead exhibiting a slope equal to that at negative voltages. This suggests that electroosmotic flow inside the pore is reversed, and the pore wall undergoes nearly complete charge inversion. Finally, for the higher scan rate of 4 V/s, charge inversion disappears in the I-V curve, and the pore shows clear ICR.

At a low scan rate of 0.02 V/s, the response time required for charge inversion is extremely long, ~38.2 s (red curve in Fig. 5c), and the amount of charge is very high, at ~446.1 nC (Supplementary Fig. 14). This is due to the slow migration of $Ca^{2+}$ ions under such conditions. At 0.5 V/s, the response time decreases to ~2.6 s, a reduction of ~93%, and the required charge decreases to ~51.4 nC. As the scan rate increases further to 3 V/s, both the response time and the required charge remain almost stable.

The loop area enclosed by the I-V curves at different scan rates can be used to characterize the memristive behavior of the pore,[49] such that a larger loop area corresponds to stronger memristive behavior (the yellow region in Fig. 5b). Fig. 5d illustrates the relationship between the loop area and the scan rate under different voltages. As the scan rate increases from 0.2 to 1.2 V/s, the loop area increases from 21.7 to 57.6 nA·V, representing an increase of ~165%. This is attributed to the relatively low threshold voltage required for charge reversal inside the pore during this process. When the scan rate continues to increase to 3 V/s, the threshold voltage gradually decreases, leading to a corresponding reduction in the loop area.

## Discussion

In summary, we have systematically investigated the dynamic regulation of surface charge properties by $Ca^{2+}$ ions on the walls of cylindrical PET mesopores under concentration gradients. By analyzing the characteristics of EOF-induced ICR under various $CaCl_2$ concentration gradients, we experimentally determined the threshold concentration of ~50 mM for charge inversion in PET mesopores. When the concentration on the low-concentration side is below the threshold (e.g., 10 mM) while that on the high-concentration side exceeds the threshold (≥100 mM), $Ca^{2+}$ ions become overadsorbed on local regions of the pore wall, inducing local charge inversion and forming bipolar charge distributions along the pore axis. This non-uniform surface charge distribution periodically reverses the EOF direction, thereby generating highly regular current oscillations. Quantitative analysis shows that the oscillation frequency increases nearly linearly from ~9 events per minute at 0.8 V to ~30 events per minute at 2 V, while the open-state probability of the pore decreases from ~16.5% at 0.6 V to ~1.7% at 2 V. The full width at half maximum of the current peak shortens from 352 ms at 1 V to 49 ms at 2 V, indicating faster dynamic switching at higher voltages. Under dynamic voltage scanning, the system exhibits typical memristive hysteresis, with the loop area reaching a maximum of 57.6 nA·V at a scan rate of 1.2 V/s. Our findings provide a quantitative description for understanding the critical concentration of charge inversion and the dynamic, non-uniform surface charge regulation induced by divalent ions, and establish a simple, convenient, and controllable platform for constructing ionic oscillators.

## Methods

### Pore fabrication

Single cylindrical pores were fabricated on poly(ethylene terephthalate) (PET) film by the track-etching technique.[59] Briefly, PET films with a diameter of 30 mm and a thickness of 12 μm were first irradiated with single energetic heavy ions at the Helmholtz Center for Heavy Ion Research in Darmstadt, Germany.[60] The irradiation step led to the formation of a latent track, a local zone of the material that is susceptible to chemical etching. The irradiated film subsequently subjected to 365-nm UV light (UVP UVGL-25, CA, USA) on both sides for 1 hour per side. Then, the film was chemically etched in 2 M NaOH solutions at 60 °C using a water bath. The etching rate of PET was ~15 nm/min,[61] and pores with target diameters were obtained by controlling the etching time. After etching, the pore diameter was characterized by the ionic conductance measurement in 1 M KCl, where the high salt concentration effectively screens the surface charges of pore walls.[14]

### Solution preparation

All chemicals used in the experiments, including potassium chloride (KCl), calcium chloride ($CaCl_2$), sodium hydroxide (NaOH), and Trizma base, were purchased from Sigma-Aldrich. Deionized water used for preparing aqueous solutions was produced by a Direct-Q 3UV system (Millipore Sigma, Burlington, MA, USA). KCl or $CaCl_2$ solutions of different concentrations were buffered with 10 mM Tris base to pH = 8. A pH meter (Mettler Toledo FE 28, China) and a conductivity meter (Mettler Toledo FE 38, China) were used to measure the pH value and conductivity of each solution, respectively. The measured electrical conductivities of the KCl and $CaCl_2$ solutions are listed in Supplementary Tab. 2.

## Data acquisition

The current-voltage (I-V) curves of the pores were recorded using a Keithley 6487 picoammeter (Keithley Instruments, Solon, Ohio, USA) and a CHI 760E electrochemical workstation (CH Instruments, Shanghai, China). In the voltage scan using the Keithley 6487, a step of 0.2 V and a delay of 1 s were adopted. While the voltage scan rates ranged from 0.01 to 10 V/s with the CHI 760E. Please note that here the electrochemical workstation provides a versatile device for the recording of I-V curves and current-time (I-t) curves. For I-t recordings, a sampling frequency of 1 kHz was used. To record the current data with a high sampling frequency of 10 kHz, the Axopatch 200B and Digidata 1550B (Molecular Devices Inc., USA) were applied under different voltages. Data analysis was performed using Clampfit 11.0.

## Finite element simulations

Theoretical modeling was developed by numerically solving Poisson-Nernst-Planck (PNP) and Navier-Stokes (NS) equations with COMSOL Multiphysics.[18, 19, 20] The mesopore was situated between two aqueous reservoirs with a width of 5 μm and a length of 5 μm. Because concentrated divalent $CaCl_2$ solutions cannot be applied in the simulations based on classical mean-field theories,[37] monovalent electrolyte KCl with a concentration gradient of 600 mM: 20 mM was applied across the pore. The diffusion coefficients for $K^+$ and $Cl^-$ ions were set to $1.96 \times 10^{-9}$ $m^2$/s and $2.03 \times 10^{-9}$ $m^2$/s, respectively.[62] For the dimension of the pore, a length of 12 μm and a diameter of 250 nm were applied, which closely correspond to the values in experimental studies. To meet the simulation objectives, different surface charge conditions of the pore wall were systematically considered, including uniformly negative surfaces, uniformly positive surfaces, and bipolar surfaces with charged segments of different lengths.[63] The surface charge density was set to ±0.04 C/m$^2$.[64, 65] The dielectric constant of water was defined as 80, and the system temperature was maintained at 298 K. A voltage of 1 V was applied across the boundaries of the liquid reservoirs. Simulation details are provided in the Supplementary Information.

## Data availability

Source data are provided with this paper. All the raw data relevant to the study are available from the corresponding author upon request.

## Acknowledgments

This research was supported by the National Natural Science Foundation of China (52105579 for Y.Q. and 52275207 for T.S), the Shandong Provincial Natural Science Foundation (ZR2024ME176 for Y.Q.), the National Key R&D Program of China (2023YFF0717105 for Y.Q.), the Basic and Applied Basic Research Foundation of Guangdong Province (2025A1515010126 for Y.Q.), and the Shenzhen Science and Technology Program (JCYJ20240813101159005 for Y.Q.).

## Author contributions

H.Z. fabricated all pores, performed experiments, and wrote the manuscript. Y.Z. performed experiments and numerical modeling. Z.G. analyzed the experimental data and wrote the manuscript. C.L. analyzed the experimental data and wrote the manuscript. T.S. analyzed the experimental data. Z.S. analyzed the experimental data and wrote the manuscript. Y.Q. designed the experiments, analyzed the experimental data, and wrote the manuscript.

## Ethics declarations

### Competing interests

The authors declare no competing interests.

## Additional information

See supplementary material for simulation details and additional simulation results.